\documentclass[12pt]{article}
\usepackage{graphicx}
\usepackage{amssymb,amsmath,amsfonts,palatino,amsthm}
\usepackage{amssymb}
\usepackage{epstopdf}
\usepackage{slashed} 
\newcommand{\f}{\begin{equation}}
\newcommand{\ff}{\end{equation}}

\begin{document}

\title{Dynamics of the cosmological and Newton's constant \\}
\author{Lee Smolin\thanks{lsmolin@perimeterinstitute.ca} 
\\
\\
Perimeter Institute for Theoretical Physics,\\
31 Caroline Street North, Waterloo, Ontario N2J 2Y5, Canada}
\date{\today}
\maketitle

\begin{abstract}

A modification of general relativity is presented in which Newton's constant, $G$ and the cosmological constant, 
$\Lambda$,  become a conjugate pair of dynamical variables.

\end{abstract}

\tableofcontents

\newpage

\section{Introduction}

In this paper a modification of general relativity is presented in which Newton's constant, $G$, and the cosmological constant, 
$\Lambda$, become a conjugate pair of dynamical variables.   They are not, however fields, but functions of a globally defined time, $t$.
This proposal is then well defined only in the presence of a condition that gauge fixes the many fingered time or refoliation gauge invariance of general relativity, giving a preferred global time.  This starts out as a gauge fixing but becomes physically meaningful when we make $G$ and $\Lambda$ functions of it.  

One way to accomplish this, within a consistent modification of the field equations, is  by using the viewpoint of shape dynamics\cite{SD1,SD2,SDreview}.  That is a recently proposed reformulation of general relativity, which is locally, but not necessarily globally, equivalent to Einstein's theory.  It features a preferred time slicing but, rather than just breaking spacetime diffeomorphism invariance, shape dynamics trades the gauge symmetry of many fingered time, or refoliation invariance, for a local three dimensional conformal invariance.  This assures that the physical degrees of freedom, the linearlized approximation and the Newtonian limit are all unchanged.

The theory then trades spacetime diffeomorphism invariance for the  gauge group given by the product of three factors, based on a $3+1$ splitting of spacetime, ${\cal M} = \Sigma_3 \times R$:

\begin{enumerate}

\item{}Diffeomorphisms of the spatial slices, $\Sigma_3 (t)$, denoted $Diff (\Sigma_3 ) $

\item{}Reparametrizations of the global time coordinate, $Diff (R )$.
\f
t \rightarrow f(t)
\ff

\item{}Volume preserving local conformal transformations on $\Sigma_3$, denoted $C(\Sigma_3 )$.  These act on the spatial metric as 
\f
q_{ab} \rightarrow e^{2 \Phi } q_{ab}
\ff
subject to the limitation that the volume of $\Sigma$ 
\f
V= \int_\Sigma \sqrt{-g} 
\ff
is unchanged.

\end{enumerate}

These have precisely four degrees of freedom per point.  So we can trade
\f
Diff ({\cal M}_4 ) \leftrightarrow Diff (\Sigma_3 )  \times C(\Sigma_3 ) \times Diff (R )
\ff

The existence of shape dynamics as a locally equivalent reformulation of general relativity invites us to consider a new class of modified gravity theories which make use of the preferred time slicing.  These theories will contain novel phenomena while preserving the linearized and Newtonian approximations.  In particular having a preferred time available allows us to contemplate making physical parameters into global dynamical variables which depend only on $t$.   There are several reasons to study this kind of hypothesis.  

\begin{itemize}

\item{}To understand why the parameters have the values they do.  A very general and powerful argument leads to the conclusion that the best way to understand why the laws in general, and the parameters specifically, are what we find them to be, would be if that they have evolved according to some dynamical principle\cite{evolve,LOTC,SU,TR,TN}. 
The idea that the constants of nature evolve on cosmological time scales is proposed from different points of view in \cite{val,Ari} and observational limits to a variation of $G$ are presented in \cite{Gdotlimits,testing}.  We will see in particular how making the cosmological constant a dynamical variable transforms the puzzle of its tiny  value, opening up new possibilities for that puzzle's resolution.  

\item{}To understand the origin of the various arrows of time.  As hypothesized by Penrose\cite{Penrose}, one strategy for explaining the arrows of time is to posit that the fundamental laws are irreversible.   

\item{}To offer new hypotheses for the solution of the puzzles of cosmology.  

\end{itemize}

A large class of modified gravity theories based on this strategy were introduced in \cite{MHL}.  Here we propose a different way to do this based on a very simple idea: make $\Lambda$ and $G$ functions of a global time, $t$. To do this in a way that does not disrupt the consistency of the field equations, we will make them a conjugate pair.   

This proposal was initially inspired by a very intriguing suggestion of Kaloper and Padilla\cite{sequestering}.  In their work they related the values of Newton's constant and the cosmological constant to averages over the lifetime of the universe of certain quantities.  This is elegant, but commits us to knowledge of the whole future history of the universe.  Here we weaken their idea so that the time derivatives of these constants are related only to averages over the spatial universe at a fixed time.  You can say that the theory proposed here is a differential version of the Kaloper and Padilla theory.  

We do this in a way which preserves reparametrizations of the global time, and we find equations for the time variation of the constants\footnote{Here and below we set units such that $c=1$.},
\begin{eqnarray}
\dot{G} &=& - \frac{2}{G_0 \mu} V
\\
\dot{\Lambda} &=& \frac{1}{G_0 \mu}    {\cal K}^{matter}  = - \frac{1}{2V} \dot{G}   {\cal K}^{matter} 
\end{eqnarray}

where $G_0$ is the present value of $G$,
\f
 {\cal K}^{matter}    =  \frac{\delta S^{matter}}{\delta G} 
\ff
$\mu$ is an adjustable constant, which can be tuned to make these predictions fit within present limits.
We note that these equations are not symmetric under time reversal, giving another example of an extension of general relativity that is not time reversal invariant\cite{MHL}.

One can ask if these relations follow necessarily from the idea that $G$ and $\Lambda$ are dynamical variables.  This does appear to be the simplest version of the idea which preserves time reparametrization invariance.   But there are other versions which break the symmetry of time reparametrizations.  In section 6 we describe one which yields tight results,
\f
\frac{\dot{G}}{G}= -3 \frac{\dot{a}}{a}
\label{Gdot4}
\ff
\f
\Lambda = G < {\cal L}^{matter}   > 
\ff
Unfortunately, (\ref{Gdot4}) disagrees with observational bounds on $\frac{\dot{G}}{G}$ by several orders of magnitude\cite{Gdotlimits}.

In the next section we introduce the theory we have been describing, whose features we illustrate in section 3 by working out the $FRW$ cosmological models.   Section 4 is devoted to the important issue of consistency of the field equations in both lagrangian and hamiltonian form, while in section 5 we visit briefly the implications of treating $G$ and $\Lambda$ as quantum variables.  Brief conclusions are in section 7. In an appendix we work out the dependence on varying $G$ of the contributions to the Hamiltonian constraint arising from gauge, spinor and scalar fields, to show that a weak form of the equivalence principle is maintained.

\section{Making dimensional parameters into dynamical variables}

We consider a non-local action which breaks the spacetime diffeomorphism invariance by the imposition of preferred spatial slices.  This is done by the imposition of a preferred spatial slicing, or decomposition ${\cal M}^4 = \Sigma \times R$.  This slicing follows a preferred time coordinate, $t$.   This breaks the spacetime diffeomorphism group down into a product of spatial diffeomorphisms and temporal reparameterizations.
\f
{Diff} {\cal M}^4 \rightarrow {Diff} \Sigma \times {Diff} R
\ff
This will be manifested by the coupling constants, $G$ and $\Lambda$ becoming functions of time.

We start with the standard  action of general relativity\footnote{We use $(-,+,+,+)$ signature and conventions of 
\cite{Wald}.}
\f
S= \int dt  \int_\Sigma d^3x \sqrt{-g} \left ( \frac{1}{G_0} ( R- 2 \Lambda ) +  {\cal L}^{matter}  \right ) 
\label{S1}
\ff
Here $\Sigma$ is a compact spatial three manifold, and $G_0$ is $4 \pi$ times Newton's constant.

We alter this action in  three steps:

\begin{enumerate}

\item{} We make the cosmological constant a function of the global time 
\f
\Lambda \rightarrow \Lambda (t)
\ff

\item{}We introduce a time dependent Newton's coupling by scaling the metric in the matter action, but nowhere else.  
We do  this by scaling the lapse\footnote{An alternative is to introduce $G(t)$ via conformal scaling,
$
g_{ab}  \rightarrow  \frac{G(t) }{G_0} g_{ab}.
$
but this would result in no coupling to scale invariant matter.}
\f
g_{00}  \rightarrow  \left ( \frac{G(t)}{G_0} \right )^2 g_{00},  \ \ \ \  g_{ij}  \rightarrow g_{ij},  \ \ \ \  g_{i0}  \rightarrow g_{i0} , \ \ \ 
\label{scaling}
\ff
As a result, the following condition holds,
\f
{\cal K}^{matter} \equiv  \frac{\delta S^{matter}}{\delta G} =  \frac{2}{G} \int_\Sigma d^3 z 
g_{00} (z) \frac{\delta S^{matter}}{\delta g_{00}(z) }=- \frac{2}{G} \int_\Sigma d^3 z 
g_{00} (z) 
T^{00} 
\label{cond1}
\ff
\item{}We add a term to the action to make $\Lambda$ and $G$ into a  canonical pair.
\f
\int dt  \Lambda \frac{\dot{G}}{G_0} \mu
\ff
\end{enumerate}

The result is the action
\f
S= \int dt  \int_\Sigma d^3x \sqrt{-g} \left ( \frac{1}{G_0} ( R- 2 \Lambda ) + \frac{G}{G_0} {\cal L}^{matter}  \right ) 
 + \int dt  \Lambda \frac{\dot{G}}{G_0} \mu
\label{S2}
\ff
Here  $\Lambda$ and $G$ are functions only of the global time.  So this action defends on a preferred slicing of spacetime as in shape dynamics.  We have scaled out of ${\cal L}^{matter} $ a dominant factor of $\frac{G(t) }{G_0} $, there may be additional dependence on $\frac{G(t) }{G_0} $ hidden in ${\cal L}^{matter} $.

 $\mu$ is a constant with dimensions of $mass \cdot l^3$.  We may note that the new term is  invariant under reparameterizations of time.  $G_0$ is the value of $G(t)$ at some fixed time.   We see that we can identify $G(t)$ and $\Lambda (t) $ as canonically conjugate quantities. 
 
 In the appendix we show that the effect of the lapse scaling (\ref{scaling}) in the Hamiltonian formulation is, for scalar, chiral spinor and gauge fields, exactly to multiply the corresponding matter contributions to the Hamiltonian constraint by $\frac{G}{G_0}$.  This is not surprising as that is after all the effect of scaling the lapse that multiplies the matter term in the Hamiltonian constraint.
We can say then that the equivalence principle is satisfied in the weak sense that all matter degrees of freedom propagate according to the same $G$ and the same four metric.  But gravitational waves propagate via a different metric.  

To make sense of the quantities $G(t)$ and $\Lambda (t)$ we have to supplement the action with a gauge condition that fixes the refoliation or many fingered time gauge invariance.  We choose the constant mean curvature gauge condition,
\f
{\cal S}(\rho ) = \int_\Sigma \rho ( \pi - \sqrt{q}  <\pi >  )
\ff
because it generates its own gauge invariance\cite{SD1,SD2,SDreview}, which is local scale transformations.
This puts these  considerations into the domain of shape dynamics.  But other gauge fixings may serve as well.

The field equations are
\begin{eqnarray}
R_{ab} - \frac{1}{2} g_{ab} R + \Lambda g_{ab}    &=& G T_{ab} 
\label{EE}
\\
\dot{G} &=& - \frac{2 V }{ \mu} 
\label{Gdot}
\\
\dot{\Lambda} &=& \frac{1}{ \mu}    {\cal K}^{matter}  = - \dot{G}   \frac{{\cal K}^{matter}} {2 V}
\label{Lambdadot}
\end{eqnarray}

We see that the equations for $\dot{G}$ and $\dot{\Lambda}$ are first order in time and so are not invariant under time reversal. 
We can also note that $\mu$ can have either sign but that, in either case, $\dot{G}$ and $\dot{\Lambda}$ will have opposite sign.  This is forced on us by the conjugate relation of $G$ and $\Lambda$ together with the normalization that positive 
${\cal K}^{matter}$ corresponds to positive energy density.
 

We have,
\f
\frac{\dot{G}}{G}= - \frac{2V }{ G \mu}
\ff
$\mu$ can be set to conform to the present observable limits
present limits \cite{Gdotlimits}
\f
\frac{\dot{G}}{G} < 10^{-13} \frac{1}{years}
\ff 
We can parametrize $\mu$ in terms of a dimensionless number $Z$ as
\f
\mu = Z \hbar R^2
\ff
where $R^{-2}=\Lambda_0$ is the present cosmological constant.  Then we have
\f
Z > 10^{120 } 
\ff
This is consistent with a conservative bound on $\dot{\Lambda}$ given by
\f
\Lambda H > \dot{\Lambda} = \frac{V}{ \mu}  <  {\cal L}^{matter} > = \frac{M_{U}}{Z \hbar R^2 }
\ff
or 
\f
Z > 10^{80} \frac{R}{\lambda_{proton}} \approx 10^{120}.
\ff

Finally, we notice that if there is no matter our theory reduces to vacuum 
general relativity with a constant cosmological constant.

\section{FRW cosmology}

To understand how the new theory differs from standard general relativity we go right away to the simple $FRW$ cosmological models. 
The reduction is defined in the Hamiltonian formulation\footnote{For details of the reduction see \cite{MHL}.} by specializing the metric to the form
\f
g_{ab}= a^2 (t) q^0_{ab}
\ff
in terms of a fixed reference metric $q^0_{ab}$, 
while the canonical momentum is restricted to
\f
\tilde{\pi}^{ab} =  \frac{1}{3a} \sqrt{q^0} q^{ab}_0  \pi (t)
\ff

The action reduces to
\f
S=  \int dt \left [ \Lambda \frac{\dot{G}}{G_0} \mu + v_0 \left (
\pi \dot{a} - N {\cal C}
\right ) \right ] 
\label{FRWaction}
\ff
where the fiducial volume of the universe is
\f
v_0 = \int_\Sigma \sqrt{q^0}
\ff

The Hamiltonian constraint, with the homogeneous lapse $N$ 
generates time reparametrizations
\f
{\cal C} = \frac{G_0}{2a}   \pi^2  - a^3 V(a)
\ff

The standard potential $V$ is
\f
V=  \frac{\Lambda}{6G_0} - \frac{k}{2 G_0 a^2} + \frac{4 \pi G \rho_0 }{3G_0 a^3 } 
\ff
We note that the $CMC$ gauge condition is  satisfied in this  case since the momenta are constant densities.  This means that $a$ and $\pi$ are invariant under volume preserving conformal transformations generated by $\cal S$.


The new equations of motion are
\f
\dot{G} = - \frac{NV }{6 \mu}
\label{Gdot2}
\ff
where $V= v_0 a^3$ is the volume of the universe.
\f
\dot{\Lambda}= \frac{4 \pi}{3} \frac{Nv_0 \rho_0}{\mu}
\label{Ldot2}
\ff
We note that (\ref{Ldot2}) is a simple linear relation, which for either sign of $\mu$ has the opposite sign of (\ref{Gdot2}).  

We vary next  by $\pi$ to find
\f
\frac{1}{N } \dot{a}= G_0  \frac{\pi}{a}  
\label{aeom}
\ff
This gives us
\f
\pi =  \frac{a^2}{N G_0 }H 
\label{pieq}
\ff
in terms of the usual Hubble constant, $H=\frac{\dot{a}}{a}$.

If we vary the action by the lagrange multiplier (or lapse), $N$ we find the 
Friedmann equation from ${\cal H}=0$, or 
\f
{\cal H}=  a^3 \left ( \frac{1}{2N^2 G_0 } H^2  - V 
\right ) =0 
\label{C}
\ff
while varying by $a$ gives an equation for $\dot{\pi}$ 
\f
\frac{1}{N} \dot{\pi} =  \frac{G_0 \pi^2}{2a^2} + 3 a^2 V- a^3 V^\prime 
\label{pieom}
\ff

Combining everything, and fixing the lapse, $N=1$, we find
the modified Friedmann equation
\begin{eqnarray}
\left ( \frac{\dot{a}}{a}  \right )^2 &=& 2G_0 V 
\nonumber
\\
&=& \frac{\Lambda}{3} -\frac{k}{a^2 } +\frac{8 \pi G \rho_0 }{3 G_0 a^3} 
\label{FRW1}
\end{eqnarray}

By expressing the equation for $\dot{\pi}$ in terms of $\ddot{a}$ we find the acceleration equation
\begin{eqnarray}
\frac{\ddot{a}}{a} &= &  2G_0 V -aG_0 V^\prime  
\\
&=& \frac{\Lambda}{3} -\frac{4 \pi G \rho_0 }{3 a^3}
 \label{FRW2}
\end{eqnarray}
We can also compute $\ddot{a}$ directly from (\ref{FRW1}).  When we do this we get a consistency relation
\begin{eqnarray}
0= \frac{dV}{dt} &=& \frac{\partial V}{\partial \Lambda} \dot{\Lambda} +  \frac{\partial V}{\partial G} \dot{G} 
\nonumber \\
&=&  \frac{N G_0 v_0 a^3 }{\mu } \left ( \frac{\partial V}{\partial \Lambda}  \frac{\partial V}{\partial G} -  \frac{\partial V}{\partial G} \frac{\partial V}{\partial \Lambda}  
\right ) =0 
\end{eqnarray}
which we see is automatically satisfied.  Hence, the equations of $FRW$ cosmology are the same as in the standard case, with the addition of the time dependence of $G$ and $\Lambda$.  Because these dependences are linked by their being conjugate  variables, $\rho_0$ is a constant as in the usual case.

\section{Consistency relations}

We just saw that the reduction of our theory to $FRW$ cosmological models is consistent, without the need for additional interactions coupling the matter energy density to the time derivatives of $G$ and $\Lambda$.  
This was due to $G$ and $\Lambda$ being conjugate variables.

\subsection{Consistency of the field equations}

We also have to check whether the equations of motion of the full theory are consistent.  We then take the covariant divergence of the Einstein equation (\ref{EE}) which, using the Bianchi identities on the $LHS$ gives us,
\f
0 = - g^{ab} \partial_a \Lambda + T_{matter}^{ab} \partial_b G +G \nabla_b T_{matter}^{ab}  
\ff
which gives us  
\f
\boxed{
\nabla_b T_{matter}^{ib}  =- \frac{\dot{G}}{G} T_{matter}^{i0} }
\label{consistenti}
\ff
when $a=i$.  The time component also gives a new local relation
\f
\boxed{
\dot{\Lambda}= \dot{G} T^{00}_{matter} + G \nabla_a T^{a0}_{matter}}
\label{consistent0}
\ff
Both (\ref{consistenti}) and (\ref{consistent0}) impose conditions on the matter degrees of freedom.  This is not hard to understand; changes in $\Lambda$ modify the vacuum energy density, which requires taking energy from or giving energy to the matter degrees of freedom. However, we saw that in the $FRW$  case this was compensated by changes in $G$, so no energy had to be requisitioned from the matter.  To what extent does this occur in the full theory?

To investigate this we integrate (\ref{consistent0}) over $\int_\Sigma \sqrt{q} g_{00} $.  This gives
\f
\dot{\Lambda}= \dot{G} <T^{00} g_{00} >+ \frac{G}{V} \frac{\partial}{\partial t }  \int_\Sigma \sqrt{q} T^{00} g-{00}
\label{consisten4}
\ff
We can compare this with 
the equation of motion (\ref{Lambdadot}).
\f
\dot{\Lambda} = -  \dot{G} \frac{ {\cal K}^{matter} }{2V}
\ff
We find
\f
\frac{\partial}{\partial t }  \int_\Sigma \sqrt{q} T^{00}_{matter} =  \frac{\dot{G}}{G} ( \int_\Sigma \sqrt{q}   g_{00}T^{00}_{matter} + 
\frac{1}{2}{\cal K}^{matter}  V ) =0 
\label{consistency2}
\ff
where we have used the condition (\ref{cond1}) in the last step.

Thus, when (\ref{cond1}) is satisfied we have that
\f
E= \int_\Sigma  \sqrt{q} T^{00}_{matter} 
\label{Edef}
\ff
is conserved in time, i.e.
\f
\frac{dE}{dt}=\frac{d}{dt} \int_\Sigma  \sqrt{q} T^{00}=0
\label{Edot}
\ff
This is a covariant conservation law because of the density factor.

\subsection{Consequences of the modified gauge invariance}

We can understand the consistency relation (\ref{consistency2}) as a consequence of the weakened diffeomorphism invariance.

We can see this by writing
\f
S=S^0 + \int dt \left [ L(t) + \mu \Lambda \frac{\dot{G}}{G}
\right ]
\ff
where $S^0$ is the pure gravitational action, which depends on $G_0$ but not on $G$ or $\Lambda$.  Hence the dependence of the field equations of $G$ and $\Lambda$ is in
\f
L(t) = \int_\Sigma d^3x {\cal L } =   \int_\Sigma d^3x \sqrt{-g} \left ( -2 \frac{\Lambda}{G_0}   + \frac{G}{G_0} {\cal L}^{matter}  \right ) 
\ff
We have under $\delta g_{ab} = {\cal L}_v g_{ab}$
\f
0 = \int dt \delta L = \int_\Sigma d^3x \frac{\delta  {\cal L }  }{\delta g_{ab}} {\cal L}_v g_{ab} 
\ff
The gauge symmetry of ${Diff} \Sigma \times {Diff R} $ is represented by $v^a = v^i (x)$ which generates spatial diffeomorphisms and $v^a = (r(t), 0,0,0)$,  which generates global time reparametrizations.  Under the first we have
\f
\partial_a \tilde{T}^{ai} = \partial_a \left (  G \sqrt{-g} {T}^{ai}_{matter} \right ) = 0
\ff
where the dentistized $\tilde{T}^{ab}$ is given by,
\f
\tilde{T}^{ab} =  \frac{\delta  {\cal L }  }{\delta g_{ab}} 
\ff
and the matter energy-momentum tensor is
\f
{T}^{ab}_{matter} =  \frac{1}{\sqrt{-g}}\frac{\delta  {\cal L }^{matter}  }{\delta g_{ab}} 
\ff
We have 
\f
\tilde{T}^{ab} = \sqrt{g} \left ( {T}^{ab}_{matter} - \Lambda g^{ab}
\right )
\ff

The spatial components give (\ref{consistenti}).

The time component is just a global relation
\f
0= \int_\Sigma \partial_a \tilde{T}^{0a} =  \int_\Sigma \partial_0 \tilde{T}^{00} 
\ff

We can write this as,
\f
\dot{\Lambda}- \dot{G} <T^{00}_{matter}  > = G < \nabla_b  T^{0b}_{matter}  >
\label{consistent4}
\ff
This is the same as the integral of the divergence of the time component of the equations of motion
\f
\dot{\Lambda}= \dot{G} <T^{00}_{matter} >+ \frac{G}{V} \frac{\partial}{\partial t }  \int_\Sigma \sqrt{q} T^{00}_{matter}
\label{consisten3}
\ff
Hence we recover the covariant conservation law (\ref{Edot}).

\subsection*{Consequences of the weakened differ invariance, take two}

To focus on this result consider the variation of $S$ under global reparametrizations of $t$, i.e.
\f
\delta t = f^\prime (t)
\ff
We have
\begin{eqnarray}
0= \delta S &=& \int dt \left [ \int_\Sigma d^3 x ( \frac{\delta L^0}{\delta g_{ab}(x)} \delta g_{ab} (x) + \frac{\delta L}{\delta g_{ab}(x) } \delta g_{ab} (x) )
+ \frac{\delta L}{\delta G} \delta G + \frac{\delta L}{\delta \Lambda} \delta \Lambda 
\right ]
\nonumber \\
&& + \frac{\mu}{G_0}  \int dt \left [   \dot{G}\dot{\Lambda} - \dot{G}\dot{\Lambda}
\right ]
\end{eqnarray}
But
\f
\frac{\delta L}{\delta G} \delta G + \frac{\delta L}{\delta \Lambda} \delta \Lambda =
\frac{\delta L}{\delta G} \dot{G} + \frac{\delta L}{\delta \Lambda} \dot{\Lambda} = 
\frac{\delta L}{\delta G}  \frac{\delta L}{\delta \Lambda} - \frac{\delta L}{\delta \Lambda}\frac{\delta L}{\delta G} =0
\ff
while 
\f
\frac{\delta L^0}{\delta g_{ab}} \delta g_{ab} = \int_\Sigma G^{ab} \nabla_a v_b = - \int_\Sigma \nabla_a  G^{ab} v_b=0
\ff
Thus we have, with $v_a = ( \gamma (t), 0,0,0 )$,
\f
0= \int_\Sigma d^3 x  \frac{\delta L}{\delta g_{ab}(x)} \delta g_{ab} (x)= - \gamma \int_\Sigma d^3 x \nabla_a ( \sqrt{g} T^{a0} )
=  - \frac{d}{dt} \gamma \int_\Sigma d^3 x \sqrt{g} T^{00} 
\ff
reproducing (\ref{Edot}).

\subsection*{Consequences of the weakened differ invariance, take three}

We can understand this from  still another angle, if we write the action in Hamiltonian form as 
\f
L(t) = \int_\Sigma d^3x \left ( \tilde{\pi}^{ij} \dot{q}_{ij} - N {\cal H} - N^i {\cal D}_i
\right )
\ff 
where $N=g_{00}$ is the lapse and $ {\cal H}$ is the Hamiltonian constraint.  We note that this has a matter term 
\f
 {\cal H} =  {\cal H}^{grav} +  {\cal H}^{matter}
\ff
and that consistency with the Lagrangian formulation requires that $ {\cal H}^{matter}$ have a non-linear dependence on $G$ such that
\f
\frac{\delta {\cal H}^{matter}}{\delta G} = - \frac{1}{G} {\cal L}^{matter}
\ff
This can be checked for example by the scalar field where
\f
 {\cal L}^{matter} = - \int_\Sigma \frac{G}{G_0}  \sqrt{-g} ( \frac{1}{2} g^{ab} \partial_a \phi \partial_b \phi - V)
\ff

Then we have 
\begin{eqnarray}
0 &=& \frac{d}{dt} \int_\Sigma \tilde{T}^{00} =  \frac{d}{dt} \int_\Sigma N \sqrt{q} {\cal H} 
\nonumber \\
&=& \frac{\partial }{\partial t} \int_\Sigma N \sqrt{q} {\cal H}+  \int_\Sigma N \sqrt{q} \frac{\partial {\cal H}}{\partial G} \dot{G} 
+  \int_\Sigma N \sqrt{q} \frac{\partial {\cal H}}{\partial \Lambda} \dot{\Lambda} 
\nonumber \\
&=&  \frac{\partial }{\partial t} \int_\Sigma N \sqrt{q} {\cal H}+  \frac{G}{\mu} 
\left ( \int_\Sigma N \sqrt{q} \frac{\partial {\cal H}}{\partial G}  \int_\Sigma N \sqrt{q} \frac{\partial {\cal H}}{\partial \Lambda}
-  \int_\Sigma N \sqrt{q} \frac{\partial {\cal H}}{\partial \Lambda}  \int_\Sigma N \sqrt{q} \frac{\partial {\cal H}}{\partial G} \right )
\nonumber \\
&=&  \frac{\partial }{\partial t} \int_\Sigma N \sqrt{q} {\cal H}
\end{eqnarray}

\subsection{Consistency of the Hamiltonian constraint formulation}

Finally, we comment on the Hamiltonian constraint algebra.  We can write the constraints in a way that makes explicit their dependence on $G$ and $\Lambda$.
\f
{\cal H}(N) = {\cal H}^0(N) + G {\cal H}^{matter} (N) + \Lambda V(N)
\ff
Where $V(N) = \int_\Sigma \sqrt{q} N $.   We have
\f
\{ {\cal H}(N) , {\cal H}(M) \} = {\cal D}(v^i = q^{ij} (N \partial_j M - M \partial_j N )) + \frac{G_0}{\mu} \left [
{\cal H}^{matter} (N) V(M)     - {\cal H}^{matter} (M) V(N)
\right ]
\ff
which doesn't close on ${\cal D} (v)$  unless $N=M$ or $V(M)=V(N)=0$.
We note that the spatial diffeomorphism constraints, ${\cal D} (v)$, and the $CMC$ conditions, ${\cal S}(\rho )$ are functions of neither $G$ nor $\Lambda$ so their algebra is unchanged.   Hence the algebra of first class constraints is generated by ${\cal D}(v)$ and ${\cal H}(1)$ and have the algebra of 
$Diff (\Sigma_3 ) \times Diff (R)$.  We can add to this the local scale transformations generated by ${\cal S}(\rho )$.  Hence we can see this as a gauge theory with four gauge transformations per point, generated by ${\cal D}(v)$, ${\cal H}(1)$ and ${\cal S}(\rho )$, where the latter is gauge fixed by ${\cal H}(N)$.  

This is  consonant with the basic idea of shape dynamics which can be expressed in these terms in the following way.  Usually in Hamiltonian approaches to general relativity, one thinks that the first class algebra generated by ${\cal D} (v)$ and ${\cal H}(N)$
generates the gauge transformations.  A partial gauge fixing is given by ${\cal S}(\rho )$; this gauge fixes the ${\cal H}(N)$ where 
$<N>=0$.  This leaves unfixed ${\cal H}(1)$.  But we can turn this around and consider the gauge invariance of the theory to be the group $Diff (\Sigma_3 )  \times C(\Sigma_3 ) \times Diff (R )$  generated by ${\cal D}(v)$, ${\cal S}(\rho )$ and ${\cal H}(1)$.  This is partially gauge fixed by ${\cal H}(N)$, subject to  
$<N>=0$, which it happens also have a first class algebra with the  ${\cal D}(v)$.  However note that these generators are not functions of $\Lambda$ because $V(N)= V <N>.$  Thus we are free to make $G$ and $\Lambda$ a canonical pair of dynamical variables without disrupting the algebra of constraints and gauge fixing functions of the theory, and hence the gauge invariances.  The dynamics can be considered to be generated by ${\cal H}(1)$ which generates reparametrizations of the global time $t$, and does incorporate consistently the effects of making $\Lambda$ and $G$ dynamical functions of $t$.

\section{Quantization}

We can make some naive first comments about the quantization of our theory.  The canonical theory extends that of general relativity by elevating $G$ and $\Lambda$ to a conjugate pair of dynamical variables with Poisson brackets
\f
\{ \Lambda, G  \} = \frac{G_0}{\mu} = \frac{G_0}{Z \hbar R^2 } 
\ff
This leads to commutation relations
\f
[ \hat{\Lambda}, \hat{G} ] =  \frac{G_0}{Z  R^2 } 
\ff
and an uncertainty relation 
\f
\Delta \Lambda \Delta G  >  \frac{G_0}{4 Z  R^2 } 
\ff
or
\f
\frac{\Delta \Lambda}{\Lambda_0 } \frac{\Delta G }{G_0} >  \frac{1}{4 Z } \approx 10^{-123} 
\ff
Thus, we do not have to worry about quantum fluctuations in $G$ or $\Lambda$ when doing observational cosmology.   

\section{An alternative}

We next consider instead the following action, which has still weaker gauge symmetry as it is not invariant under reparameterizations of time.  
\f
S= \int dt \left [ \int_\Sigma d^3x \sqrt{-g} \left ( \frac{1}{G_0} ( R-2 \Lambda ) - \frac{G}{G_0} {\cal L}^{matter} \right ) 
-\frac{\Lambda}{G} R^3 
\right ]
\ff
$R$ is a new, fixed cosmological length.

The field equations are
\begin{eqnarray}
R_{ab} - \frac{1}{2} g_{ab} R &=& G T_{ab} + \Lambda g_{ab}   
\\
\frac{2}{G_0} V &=& \frac{R^3}{G}
\\
\frac{1}{G_0} \int_\Sigma \sqrt{-g} {\cal L}^{matter} &=& \Lambda \frac{R^3}{G^2}
\end{eqnarray}

In addition to the Einstein's equation we have two equations that fix $\Lambda$ and $G$ as functions of time.
\begin{eqnarray}
G &=&  G_0 \frac{R^3}{V}
\label{Geq}
\\
\Lambda &=& G < {\cal L}^{matter}   > =  G_0 \frac{R^3}{V} < {\cal L}^{matter}   >
\end{eqnarray}
where
\f
< {\cal L}^{matter}   > = \frac{\int_\Sigma \sqrt{-g} {\cal L}^{matter} }{V}
\ff

Additional relations are  imposed by the Bianchi equation.  Let us write
\f
T_{ab}= T_{ab}^0 + \tau_{ab}
\ff
where $ \nabla_b T_a^{0 b} =0 $.

Then taking the divergence of the Einstein equation we have
\f
0= g_{a}^0 \dot{\Lambda} + T_{a}^0 \dot{G} + G \nabla_b \tau_a^b 
\ff
We can break this up to space and time equations
\begin{eqnarray}
0 &=&  \dot{\Lambda} + \rho  \dot{G} + G \nabla_b \tau_0^b 
\\
0 &=& N_{i} \dot{\Lambda} + T_{i}^0 \dot{G} + G \nabla_b \tau_i^b 
\end{eqnarray}

Hence there must be a component of the matter field which is not conserved.  The second equation can be taken as determining the shift, $N_i = g_i^0$.  

The value of the cosmological  constant seems plausible.
However from  (\ref{Geq})  we can easily see that
\f
\frac{\dot{G}}{G}= -3 \frac{\dot{a}}{a}
\ff
which is of the order of the Hubble constant.  However this contradicts present limits \cite{Gdotlimits}
\f
\frac{\dot{G}}{G} < 10^{-13} \frac{1}{years}
\ff

\section{Conclusions}

It will be interesting to develop this theory and understand if it offers any insights to the puzzles of cosmology. 
Beyond that several new directions beckon, heralded by two queries:  1)  Can the other parameters in the laws be also turned into dynamical variables?  Is there a principle that tells us how to put them together into conjugate pairs?  2)  We have made two parameters dynamical, but at the cost of introducing a new constant, $\mu$.  Shouldn't $\mu$ itself be dynamical? 

We note that the theory proposed here is very constrained by observations.  $\frac{\dot{G}}{G}$ is very tightly constrained; to the same level the theory predicts violations of the equivalence principle in the strong sense that matter and gravitational waves propagate according to two metrics, which differ by the factor of $(\frac{G}{G_0})^2 $ in $g_{00}$.  But the equivalence principle is satisfied in the weak sense that all matter degrees of freedom propagate according to the same metric as shown in the appendix.  

The value required for $Z$ by observation is intriguing as $10^{120}$ reminds us of the area of the cosmological horizon in Planck units, and hence holography.  But the key questions to be addressed are whether the new hypothesis of varying $G$ and $\Lambda$, posited to be a conjugate pair, can shed any light on the early universe or the cosmological puzzles.  It will be also good to see if these ideas make contact with previous explorations of the idea of physical parameters varying on a cosmological scale such as those discussed in \cite{val,Ari,testing,cuscuton}.

\section*{ACKNOWLEDGEMENTS}

This is a step in a program of work investigating the hypothesis that time is real and irreversible, which we put forward in \cite{SU,TR,TN} and developed in \cite{ECS1,ECS2,MHL}. I would like to thank my collaborators Roberto Mangabeira Unger, Marina Cortes and Henrique Gomes as well as  Linqing Chen, Joao Magueijo and, especially, Niayesh Afshordi,  for conversations on this work.  I would also like to thank several people who made perceptive comments in a talk based on this work\cite{talk}.

This research was supported in part by Perimeter Institute for Theoretical Physics. Research at Perimeter Institute is supported by the Government of Canada through Industry Canada and by the Province of Ontario through the Ministry of Research and Innovation. This research was also partly supported by grants from NSERC, FQXi and the John Templeton Foundation.

\appendix

\section{Couplings to matter fields}

We illustrate the procedure of lapse scaling first in the case of Maxwell theory, then give the results for scalars and chiral fermions.  We start with the standard action
\begin{eqnarray}
S^{Maxwell} &=& -\frac{1}{4} \int_{\cal M} \sqrt{-g} g^{ab} g^{cd} F_{ac} F_{bd}
\\
&=& -\frac{1}{2} \int_{\cal M} \sqrt{-g_{00} } \sqrt{q} \left (
g^{00} g^{ij} F_{0i} F_{0j} - q_{ij} B^{i} B^{j}
\right )
\end{eqnarray}
We carry out (\ref{scaling}) to find, with $g_{00}= -N^2$, 

\f
S^{Maxwell} =  \int_{\cal M} \sqrt{q} \left ( \frac{G_0}{2GN}
g^{ij} F_{0i} F_{0j} - \frac{GN}{2G_0}  q_{ij} B^{i} B^{j}
\right )
\ff
We define the momenta
\f
\tilde{\pi}^i = \frac{\delta S^{Maxwell} }{\delta \dot{A}_i} = \frac{G_0\sqrt{q}}{GN}E^i
\ff
in terms of which we write
\f
S^{Maxwell} = \int_{\cal M} \left ( \tilde{\pi}^i \dot{A}_i - A_0 \partial_j \tilde{\pi}^j - N H^{Maxwell}
\right )
\ff
where the electromagnetic contribution to the Hamiltonian constraint is multiplied by $G$,
\f
H^{Maxwell} = \frac{G}{2G_0} \left (
\tilde{\pi}^i \tilde{\pi}^j \frac{q_{ij}}{\sqrt{q}} + B^i B^j q_{ij}  
\right )
\ff
The same goes for the standard actions for scalar and chiral fermion fields.
\f
S^{\phi + \Psi} =  \int_{\cal M} \sqrt{-g} \left ( -\frac{1}{2} g^{ab} \partial_a \phi \partial_b \phi - V(\phi ) 
+ \Psi^\dagger_{A'}  \sigma^{\mu A'}_A e^a_\mu ({\cal D}_a \Psi )^A \right )
\ff
We scale the lapse by (\ref{scaling}) and find the corresponding contributions to the Hamiltonian constraint, which is
\f
H^{\phi +\Psi } = \frac{G}{2G_0} \left ( \frac{1}{2\sqrt{q} } \pi^2 +  \frac{1}{2} q^{ij} \partial_i \phi \partial_j \phi + V(\phi )
+ \Pi_A \tau^{I A}_B e_I^i ({\cal D}_i \Psi )^B
\right )
\ff
where the momenta are defined by $\pi = \frac{G_0}{GN} \sqrt{q}  \dot{\phi} $ and 
$\pi^A = \sqrt{q} \Psi^\dagger_{A'} \sigma^{0A'A}$, while $\sigma^\mu_{A'A}$ are the four dimensional spin matrices, related to
the three dimensional Pauli matrices by $\tau^{I A}_B = \sigma^0_{A'A} \sigma^I_{A'B}$.   

So we see that the effect of carrying out the scaling (\ref{scaling}) on the lapse is to multiply uniformly the matter part of the Hamiltonian constraint by $\frac{G}{G_0}$.

\end{document}